\documentclass[useAMS,usenatbib]{mn2e}
\usepackage{times,aas_macros}
\input{epsf}

\title[]
{Modelling the Broadband Spectra of MCG-6-30-15 with a Relativistic Reflection Model}

\author[C.-Y. Chiang \& A. C. Fabian]
{Chia-Ying Chiang$^{1}\thanks{E-mail: cychiang@ast.cam.ac.uk}$ and A. C. Fabian$^{1}$\\
$^1$Institute of Astronomy, University of Cambridge, Madingley Road, Cambridge
CB3 0HA\\}

\date{
Accepted 2011 February 16.  Received 2011 February 7; in original form 2010 October 13}
\pagerange{\pageref{firstpage}--\pageref{lastpage}} \pubyear{2010}

\begin{document}

\topmargin = -0.5cm

\maketitle

\label{firstpage}

\begin{abstract}
The spectrum of the bright Seyfert I galaxy MCG-6-30-15 shows a broad Fe K$\alpha$ emission line,
generally interpreted to originate very close to the central black hole which must then have a high spin.
The observed X-ray variability is driven by a powerlaw component, with
little variability found in the iron line energy band.
The disconnection of continuum and line variability may be a consequence of strong gravitational light bending.
The X-ray spectrum of MCG-6-30-15 is however complex, strongly modified at soft X-ray energies by warm absorbers,
which some workers have extended to build an absorption-dominated model of the whole source behaviour.
The absorption model interprets the whole X-ray spectrum without a relativistic reflection component and
attributes most variability to the warm absorbers.
We re-examine the \emph{XMM-Newton}, \emph{Chandra} and \emph{Suzaku} data taken in 2001, 2004 and 2006, respectively, and construct a model consisting of several warm absorbers confirmed by spectral analysis, together with a relativistically blurred reflection component that explains both the soft and hard excesses as well.
The model works well on data from all epochs, demonstrating that the reflection model provides a consistent interpretation of the broadband spectrum of MCG-6-30-15.

\end{abstract}


%==============================================
\section{Introduction} \label{sec:introduction}\
%==============================================
The X-ray reflection spectrum of Active Galactic Nuclei (AGN), included fluorescent lines, the Compton hump and other emission in the low-energy band,
is expected from a cool accretion disc illuminated by a power-law continuum \citep{Ross93}.
As cold gas absorbs high energy photons, inner-shell electrons are knocked out and, followed by atomic transitions, lead to emission lines.
The iron K$\alpha$ line which lies at or above 6.4 keV is usually the strongest.
This fluorescent line is intrinsically narrow, but can appear as a broad line because of Doppler effect  and gravitational redshift effects \citep{Fabian89,Fabian00,Reynolds03}.
A broad iron line was first discovered by the Advanced Satellite for Cosmology and Astrophysics (\emph{ASCA}) in the X-ray spectrum of the
Seyfert I galaxy MCG-6-30-15 \citep{Tanaka95}.
Since then MCG-6-30-15 has become one of the most studied AGNs.
The broad and skewed iron line was observed in later observations of \emph{BeppoSAX} \citep{Guainazzi99}, \emph{Chandra} \citep{Lee02}, \emph{XMM-Newton} \citep{Fabian02} and \emph{Suzaku} \citep{Miniutti07}.

The soft X-ray spectrum of MCG-6-30-15 has complex structures usually attributed to warm absorbers \cite[e.g.][]{Otani96}.
\citet{BR01} tried however to explain the soft features by the same mechanism used to interpret the broad iron line.
They described the 0.7 keV drop as the blue wing of a broad O {\sevensize VIII} emission line, and the smaller $\sim$0.39 keV and $\sim$0.54 keV
drops as C {\sevensize VI} and N {\sevensize VII} emission lines, respectively.
Reflection modelling of the soft X-ray lines \citep{Ballantyne02} failed to confirm the large equivalent widths required by the data.
\citet{Lee01} presented evidence of dusty warm absorbers in the \emph{Chandra}
High Energy Transmission Grating Spectrometer (HETGS) observation.
The 0.7 keV drop can be interpreted as a combination of resonance absorption lines redward of the O {\sevensize VII} edge
and an Fe {\sevensize I} L-shell absorption edge, caused by dust grain iron oxides or silicates.
The presence of dust is in agreement with the optical observations of \citet{Reynolds97}.
\citet{Turner03} examined the Reflection Grating Spectrometer (RGS) spectra from \emph{XMM-Newton} observation and confirmed the presence
of dusty warm absorbers.
The total X-ray spectrum of MCG-6-30-15 could hence be illustrated by a power-law continuum, the reflected component from the cool disc,
and absorptions by warm absorbers.

An initially puzzling lack of line variations during rapid continuum variations \citep{Fabian02,Matsumoto03} aroused interest.
\citet{Miniutti04} showed that if the illuminating source moves within the innermost area around the black hole where there are strong general relativistic light-bending effects, then the flux of the continuum varies but the iron line could stay unchanged.
This gravitational light-bending model interprets the lack of variability of the iron line, despite the observed variability of the continuum, as due to changes in the location of the powerlaw component.
Further studies of this model have been reported by \citet{Suebsuwong06}, \citet{Niedzwiecki10}, and \citet{Niedzwiecki08}.
The timing analysis of the \emph{Suzaku} observation in 2006 later showed consistency with the gravitational light-bending model \citep{Miniutti07}.
\citet{Inoue03}, on the other hand, suggested a different view with an absorption-dominated model which could explain the X-ray spectra of MCG-6-30-15 as well.
The warm absorbers might imitate the ``red wing" structure seen in the iron line energy band, and variations of continuum are hence
due to warm absorbers but not of the power-law component.
\citet{Miller08,Miller09} further extended the absorption model to build a complex ``3+2" model which consists of five absorbing zones, two of which are `partial-covering'.
With 3 percent systematic errors included, they claim that this model reproduces the variability and \emph{XMM} spectra of MCG-6-30-15.
The partial covering component responsible for the red wing shape produces no other observed features at CCD resolution.
Using \emph{Suzaku} data, \citet{Miyakawa09} also favour this view but reduce the need for five warm absorbers to two.

Apart from the variability disconnection, it has been claimed that the strong emission found at high energies (above 10keV), known as the ``hard excess", cannot be explained with a simple disc reflection scenario.
\citet{Miller09} concluded that high-column partial-covering absorption dominates the hard excess in MCG-6-30-15, as seen in NGC 1365 \citep{Risaliti09},
1H 0419-577 \citep{Turner09} and PDS 456 \citep{Reeves09}.
The spectra of the last three objects have however since been shown to be well described by reflection \citep{Walton10},
although Compton thick, partial-covering absorbers could not be completely ruled out.

We re-examine the full X-ray data of MCG-6-30-15 taken after 2001 by \emph{XMM-Newton}, \emph{Suzaku} and \emph{Chandra}.
A reflection model with absorbing zones was applied to these observations to test the robustness of the model.
We start by investigating the difference spectra of the three datasets and then study the full real spectra.
The grating data of \emph{Chandra}, the energy resolution of which is better than other instruments, were used to study the effects of warm absorbers
and determine their shape.
A model with four absorbers works well on \emph{Chandra} and \emph{Suzaku} spectra, while the spectrum of \emph{XMM-Newton} needs
only three absorbers to be fitted.
Changes of ionization state in at least one absorbing zone are required to fit data taken at different epochs.
The partial-covering absorption model might offer an alternative explanation for the variability and spectrum of MCG-6-30-15, but
the reflection model, together with the light-bending model, is fully consistent with the data and provides a more physical point of view.

%========================================
\section{Data reduction} \label{sec:data}
%========================================
Data from \emph{XMM-Newton}, \emph{Suzaku} and \emph{Chandra}
provide observations of different epochs of MCG-6-30-15. These
observations are in different flux states so that we can study
spectrum variations and evolution in recent years. The 2-10 keV
bolometric flux is $4.26\times10^{-11}$ ergs cm$^{-2}$ s$^{-1}$ for
the \emph{XMM-Newton} PN, $4.14\times10^{-11}$ ergs cm$^{-2}$
s$^{-1}$ for the \emph{Suzaku} FI XIS, and $4.03\times10^{-11}$ ergs
cm$^{-2}$ s$^{-1}$ for \emph{Chandra} HEG. The simultaneous
\emph{BeppoSAX} observation with \emph{XMM-Newton} and PIN data of
\emph{Suzaku} offer hard X-ray data above 10 keV and we use them for
the study of hard excess.

\subsection{\emph{XMM-Newton} and \emph{BeppoSAX}}

We use the \emph{XMM-Newton} observation taken from 2001 July 31 to August 05.
In this observation EPIC MOS and PN cameras were operated in small window mode.
The observation data files (ODFs) were reprocessed with the standard procedures of
\emph{XMM-Newton} Science Analysis System ({\sevensize SAS V9.0.0}).
Source data were extracted by {\sevensize XSELECT} using a circle with a radius of 30 arcsec.
The {\sevensize HEASoft} package is incorporated into {\sevensize SAS}.
When saving the extracted spectrum in {\sevensize XSELECT}, the script
\texttt{xsl\underline{ }xmm\underline{ }epic\underline{ }makeresp} runs and creates the
redistribution matrix file ({\sevensize RMF}) and the ancillary response file ({\sevensize ARF}).
Background was taken from a region within the small window using a circle of the same size.
Using the \texttt{epatplot} tool in {\sevensize SAS} we confirm the
observation is not piled up.
Total exposure time of the PN camera is about 227 ks.
The effective energy band of EPIC cameras is 0.5-10 keV.

We reduced the Reflection Grating Spectrometer (RGS; 0.4-2.0 keV) data following the standard procedures.
Both the first order spectra of RGS1 and RGS2 data were used.
Data of these instruments overlap over most of their effective energies.
To analyse the data more efficiently we used RGS1 spectrum between 0.4 and 1.7 keV, and RGS2 spectrum between 0.9 and 1.2 keV.
We use only RGS data, which are with better energy resolution, at low energies in all our fittings.

To study the spectral variability, difference spectra were produced by subtracting the
low-flux state spectrum from the high-flux state one.
Flux cuts were chosen in a similar way mentioned in \citet{Vaughan04}.
We divided the EPIC PN 0.2-10 keV light curve into two flux intervals by 20 counts s$^{-1}$
so that the two slices have a comparable number of total counts ($\sim3\times10^{6}$).
The slice with the count rate lower than 20 was used as the low-flux state spectrum,
while the upper slice is devoted the high-flux state spectrum.

\emph{BeppoSAX} observed the source during the 2001 \emph{XMM-Newton} observation.
Data were reduced following the guidelines in \citet{Matt97}.
We use data from the Phoswitch Detector System (PDS; 13-200 keV) to
extend the coverage of the observation to the higher energy band.
The total exposure time of the PDS data is about 49 ks.

\subsection{\emph{Suzaku}}

\begin{figure}
\begin{center}
\leavevmode \epsfxsize=8.5cm \epsfbox{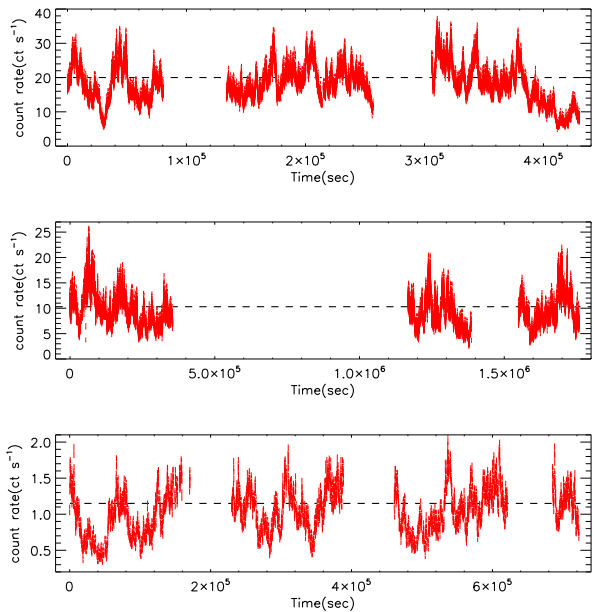}
\end{center}
\caption{The light curves on this figure are of the three satellites we use.
The top panel is the \emph{XMM-Newton} PN 0.2-10 keV light curve, while the middle one belongs to
the \emph{Suzaku} FI XIS detectors (XIS0, XIS2 and XIS3) of 0.5-12 keV,
and the bottom panel shows the combined \emph{Chandra} 0.45-10 keV light curve of first order HEG and MEG instruments.
The dash line in each panel lies on the edge of low-flux and high-flux states.
The value was chosen to make the low-flux and high-flux slices have comparable total number
of photons ($\sim3\times10^{6}$ for \emph{XMM-Newton}, $\sim1.6\times10^{6}$ for \emph{Suzaku}, and
$\sim2.6\times10^{5}$ for \emph{Chandra}).}
\label{lc}
\end{figure}

In 2006 January, MCG-6-30-15 was observed three times by \emph{Suzaku}
for a total of about 338 ks of exposure time.
The X-ray Imaging Spectrometer (XIS) was operated in the normal mode.
$5\times5$ and $3\times3$ editing modes were used throughout the observation.
We use {\sevensize HEASoft V6.8} software package  provided by \emph{NASA} to
reduce the data following the Suzaku Data Reduction Guide.
We created new cleaned event files using the \texttt{XISPI} tool.
Source products were extracted by {\sevensize XSELECT} using a circular region of
$4'.3$ radius, while background products were extracted from a smaller circular region away
from the source.
Response files were produced by the script \texttt{XISRESP}, which calls \texttt{XISRMFGEN} and
\texttt{XISSIMARFGEN} automatically.
Spectra of the three front-illuminated (FI) CCD XIS detectors (XIS0, XIS2 and XIS3) were combined
using the {\sevensize FTOOL} \texttt{addascaspec}.
For the difference spectrum, the low-flux state and high-flux state were defined in the same
way as for \emph{XMM} (see Fig. \ref{lc}).

The Hard X-ray Detector (HXD) was operated in XIS-nominal pointing mode.
The background spectrum of HXD/PIN consists of a non-X-ray background and a cosmic X-ray background.
We used the non-X-ray background event file from a database of background observations
made by the PIN diode.
The appropriate response files for the observation epoch were used as well.
The cosmic X-ray background was obtained by model estimation using the PIN response for
flat emission distribution.
We combined the source spectra and non-X-ray background spectra of different
datasets first, then created a cosmic X-ray background spectrum for the entire observation,
and added it to the total non-X-ray background spectrum.

\begin{figure}
\begin{center}
\leavevmode \epsfxsize=8.5cm \epsfbox{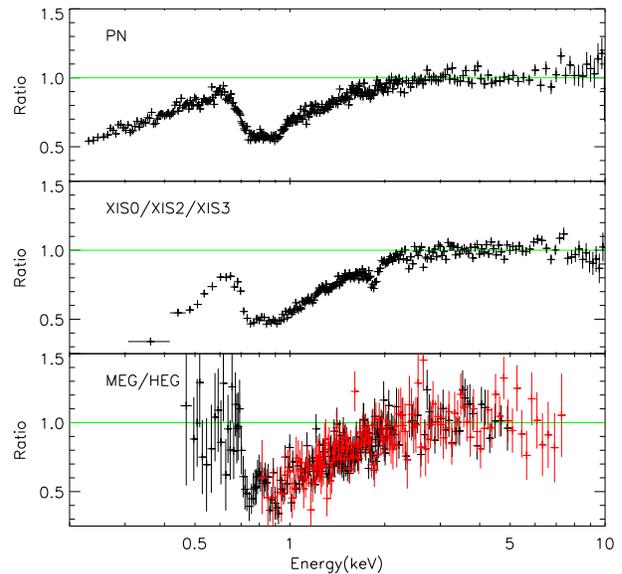}
\end{center}
\caption{The figure shows the difference spectra of the three
satellites plotted as energy against the data/model ratio. The model
of a power-law with Galactic absorption modification fitted across
3-10 keV (3-5 keV for \emph{Chandra} MEG). The difference spectra
from top to bottom belong to \emph{XMM-Newton} PN camera,
\emph{Suzaku} FI XIS detectors and \emph{Chandra} HETGS,
respectively. The absorption structure around 1.8 keV in the middle
panel is possibly caused by the detector. These difference spectra
show a highly similar structure between 0.5 keV and 5 keV. }
\label{diff}
\end{figure}

\subsection{\emph{Chandra}}

The \emph{Chandra} HETGS observed MCG-6-30-15 between 2004 May 19 and 27.
Tools in the latest {\sevensize CIAO V4.2} software package were used to reduce the data.
We started with level 1 raw data and followed standard procedures in the analysis guide
\footnote{http://cxc.harvard.edu/ciao/guides/}.
We examined the background light curve and removed periods
affected by background flaring, and obtained a good exposure time of
about 497 ks in total.
The \texttt{tgdetect} tool was used to find the zeroth and $\pm1$ order position.
With appropriate filters, a level 2 event file could be generated and a type {\sevensize II} pha (hereafter pha2) file
which contains complete information could be extracted.
Different orders of spectra of both the medium-energy gratings (MEG) and high-energy gratings
(HEG) were separated from the pha2 file.
In order to do the grating analysis in {\sevensize XSPEC}, we also split the background
file from the pha2 file by executing the \texttt{tg\underline{ }bkg} script.
{\sevensize RMF} and {\sevensize ARF} files were created by the \texttt{mkgrmf} and the
\texttt{fullgarf} tools, respectively.
Since the zeroth-order source is piled up, only $\pm1$ order spectra were used for further study.
The combined HEG (0.8-7.5 keV) and MEG (0.4-5 keV) first-order light curve is shown in Fig. \ref{lc}.\\

%==============================================
\section{Spectral Fitting} \label{sec:fitting}
%==============================================
\subsection{A Quick Look at Difference Spectra} \label{sec:difference}

We create a new spectrum by subtracting a low-flux spectrum from a high-flux spectrum, where the definition of low-flux and high-flux
states has been elaborated in Section \ref{sec:data}.
The resulting spectrum, which is called a difference spectrum, offers information on the nature of the variability because constant components in the original spectrum are removed.
The difference spectrum was initially introduced into the analysis of MCG-6-30-15 as a demonstration that the variability is indeed consisted with a pure powerlaw (as further shown here).
The absorption is a multiplicative component so in the case of a variable powerlaw, absorption components are revealed clearly in the difference spectrum (Fig. \ref{diff}).
If the variability is more complex, and no simple spectrum emerges, then a difference spectrum is difficult to interpret.
Both \citet{Fabian02} (using data of \emph{XMM-Newton}) and \citet{Miniutti07} (using data of \emph{Suzaku}) reported that
the reflection component remains constant throughout the observation period and does not appear in the difference spectra.
In addition, the energy band above 3 keV in the difference spectra can be simply fitted by a power-law with Galactic absorption
($N_{\mbox{\scriptsize H}}=4.06\times10^{20}$cm$^{-2}$), which means the warm absorbers have little effect on the spectrum above 3 keV, particularly in the Fe-K band, in contrast with the band below 3 keV.

We reproduce the difference spectra of \emph{XMM-Newton} and
\emph{Suzaku} and confirm the conclusions in previous work. The
difference spectra of \emph{Chandra} HETGS are generated to
compare changes in the difference spectra from
observations taken at difference epochs. The difference spectra of
the three satellites are shown as a ratio to a simple power-law
model with Galactic absorption in Fig. \ref{diff}. They are
remarkably similar at low energies; they are much alike below 3 keV,
at least down to 0.5 keV, which is the lowest energy we can trust
\emph{XMM-Newton} and \emph{Suzaku}. An obvious drop below 2 keV is
seen and goes to nearly the same depth in all difference spectra.
This hints that the low energy spectra we obtained can be
interpreted by the same model, and the warm absorbers did not change
much between epochs. The 3-10 keV (3-7.5 keV for \emph{Chandra})
difference spectra can be fitted by a simple powerlaw, implying the
warm absorbers cause little curvature above 3 keV. There is no clue
that warm absorbers mimic the ``red wing" structure at iron line
energies.

\begin{figure}
\begin{center}
\leavevmode \epsfxsize=8.5cm \epsfbox{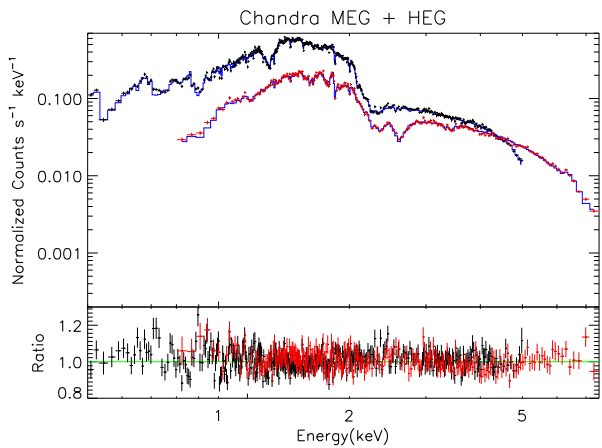}
\end{center}
\caption{The black data points in the upper panel belong to the MEG spectrum, while the red ones are of the HEG spectrum, and
the blue line stands for the best-fitting model, which consists of a powerlaw continuum, a ionised and a neutral reflection components, an additional narrow Gaussian line,
and four absorbing zones. The lower panel shows the data/model ratio of this fitting, and the data points have been re-binned for clarity.
Detailed information of this fitting is listed in Table \ref{absorber} and \ref{fitting}. The model is in excellent agreement with the data
above 1 keV and struggles slightly below 1 keV, but main features are fitted.
The small box shows the zoom-in iron line band.}
\label{chandra}
\end{figure}

\begin{figure}
\begin{center}
\leavevmode \epsfxsize=8.5cm \epsfbox{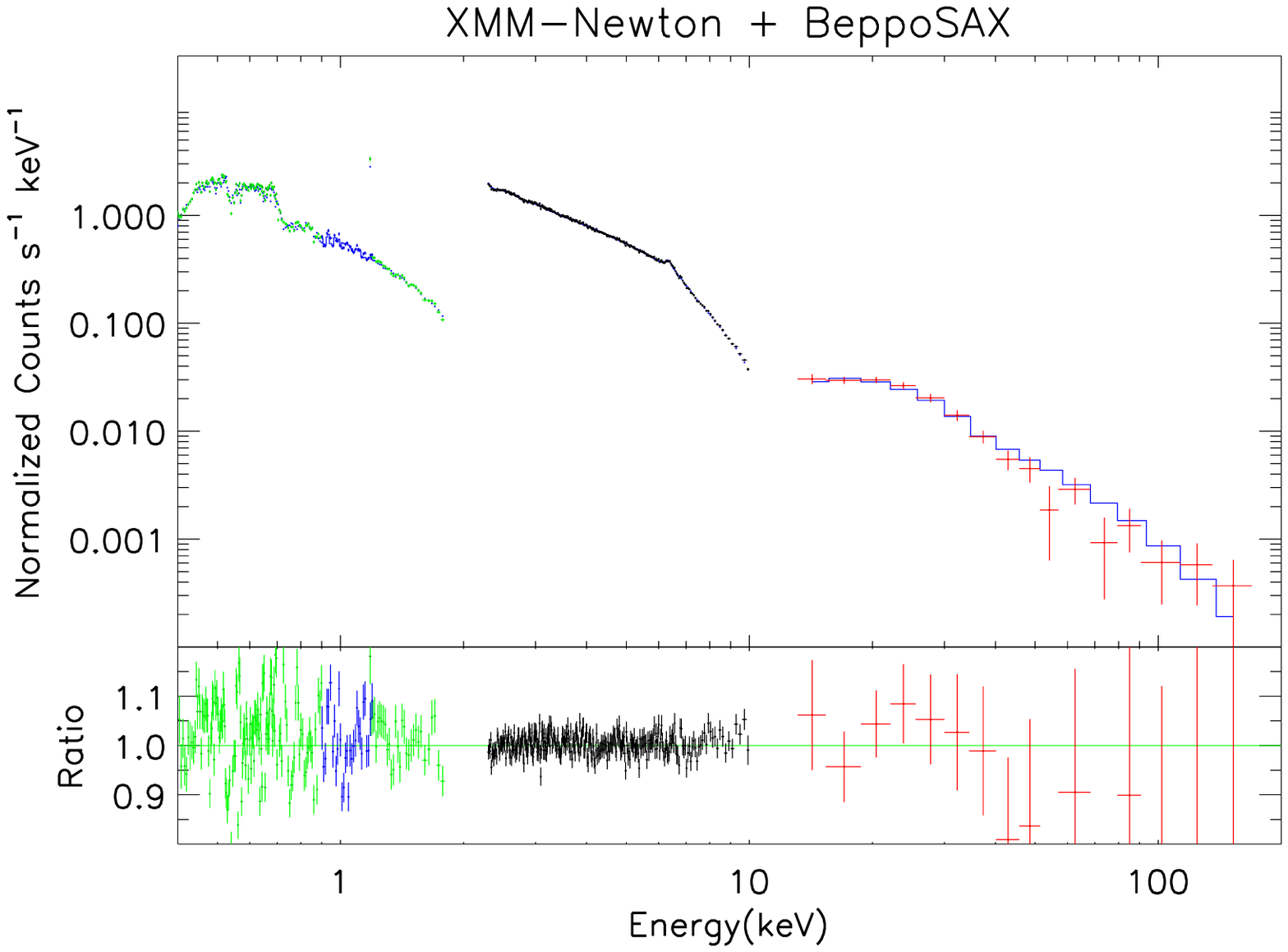}
\leavevmode \epsfxsize=8.5cm \epsfbox{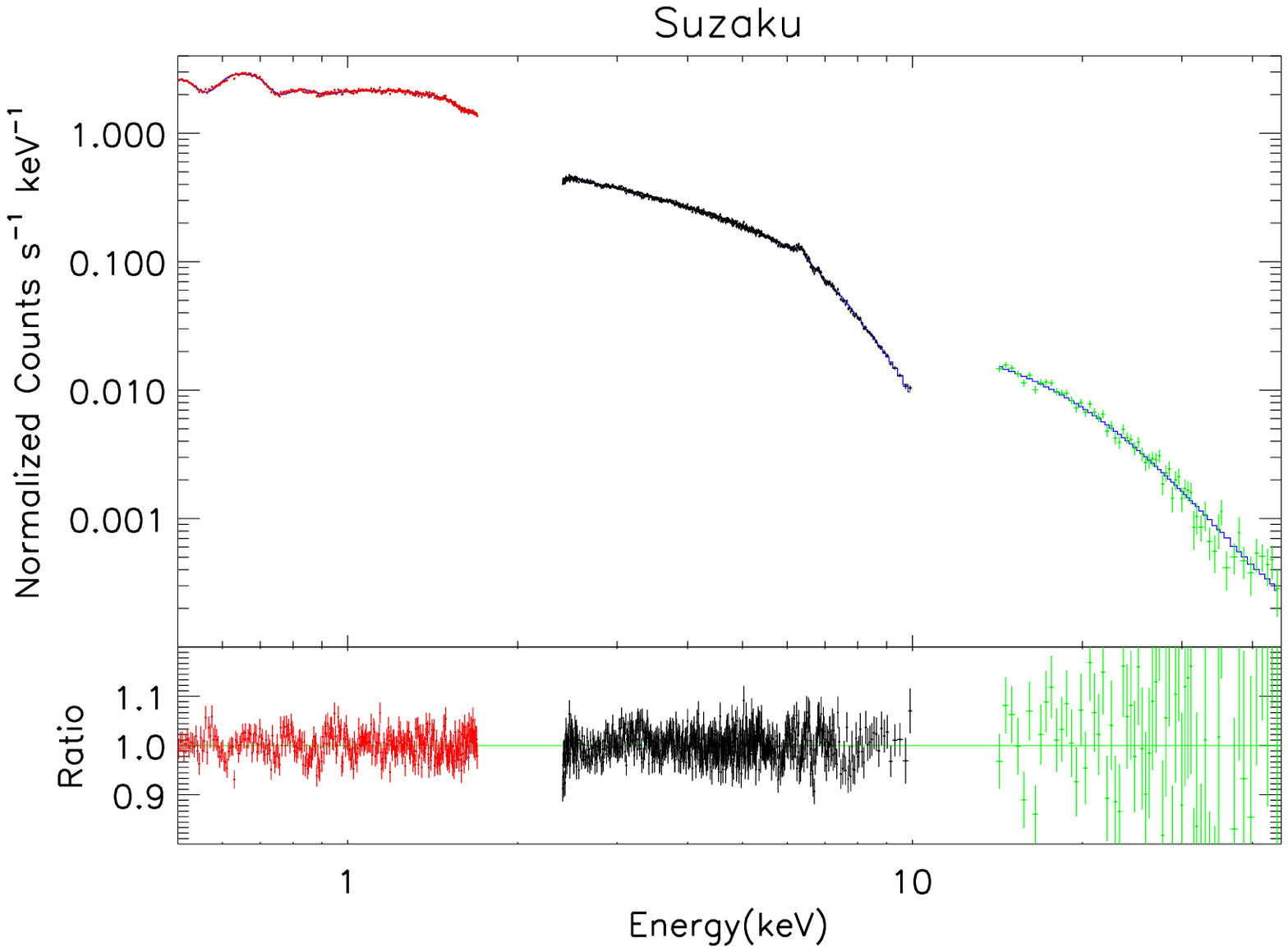}
\end{center}
\caption{The upper figure shows the best fit of the simultaneous \emph{XMM} and \emph{BeppoSAX} observation,
and the lower one of the \emph{Suzaku} observation. The black data points of \emph{XMM} in the upper figure are heavily re-binned
for clarity, and data points of \emph{BeppoSAX} are shown in red. However, the red data points in the lower figures are of
\emph{Suzaku} XIS1 detector, whose CCD is back-illuminated and collects better data at lower energies than those of front-illuminated detectors.
The high energy spectrum (above 10 keV) of \emph{Suzaku} PIN is shown in green in the lower figure.
The blue line again represents the theoretical model. It could be clearly seen that the model fits the high energy data well in both figures.
The small boxes inside the figures as well show the zoom-in iron line profile.}
\label{hard}
\end{figure}

\subsection{Warm Absorbers} \label{subsec:absorbers}

The spectrum of MCG-6-30-15 shows complicated absorption in the soft X-ray band.
\citet{Lee01} showed at least two ionisation zones are needed to interpret the absorption lines.
\citet{Sako03} verified two outflow components with velocities of -150 km s$^{-1}$ and -1900 km s$^{-1}$.
Later studies \citep[e.g.][]{Turner04,Young05} also confirmed the existence of the two kinematic components, with slightly different velocities.
\citet{McKernan07} reported both components and found two ionisation states in the slow components.
\citet{Miller08} adopted the three absorption zones mentioned in previous literature and added two more offset components to yield a good fit.
The latest detailed kinematic analysis of \citet{Holczer10}, which used the absorption measure distribution (AMD) method which
enables a full reconstruction, indicated there is a ``local" ($z$ = 0) component with a column density comparable to that of the
Galactic absorption column of this object.
The newly found third kinetic system ($\sim$-2300 km s$^{-1}$, the same shift as the redshift $z$ = 0.007749 of MCG-6-30-15) was
confused with the fast component ($\sim$ -1900 km s$^{-1}$) because of similar velocities in the local frame of reference.
\citet{Holczer10} explained the component as absorption by ionised Interstellar Medium (ISM) in the Milky Way or areas nearby.
So far the local absorption in MCG-6-30-15 is not yet confirmed by other work and its turbulence velocity remains unknown.
Summarising the previous work, a highly-ionised (log $\xi > 3.5$) absorption zone with turbulence velocity
$v_{\mbox{\scriptsize turb}}$ = 500 km s$^{-1}$, and at least two zones of $v_{\mbox{\scriptsize turb}}$ = 100 km s$^{-1}$ with
different ionisation states are required to explain the \emph{Chandra} HETGS data \citep[see Table 5 and 6 in][]{Holczer10}.

\begin{figure*}
    \begin{center}
        \begin{minipage}{1\textwidth}
            \begin{minipage}{0.33\textwidth}
                \begin{center}
                    \leavevmode \epsfxsize=50mm \epsfbox{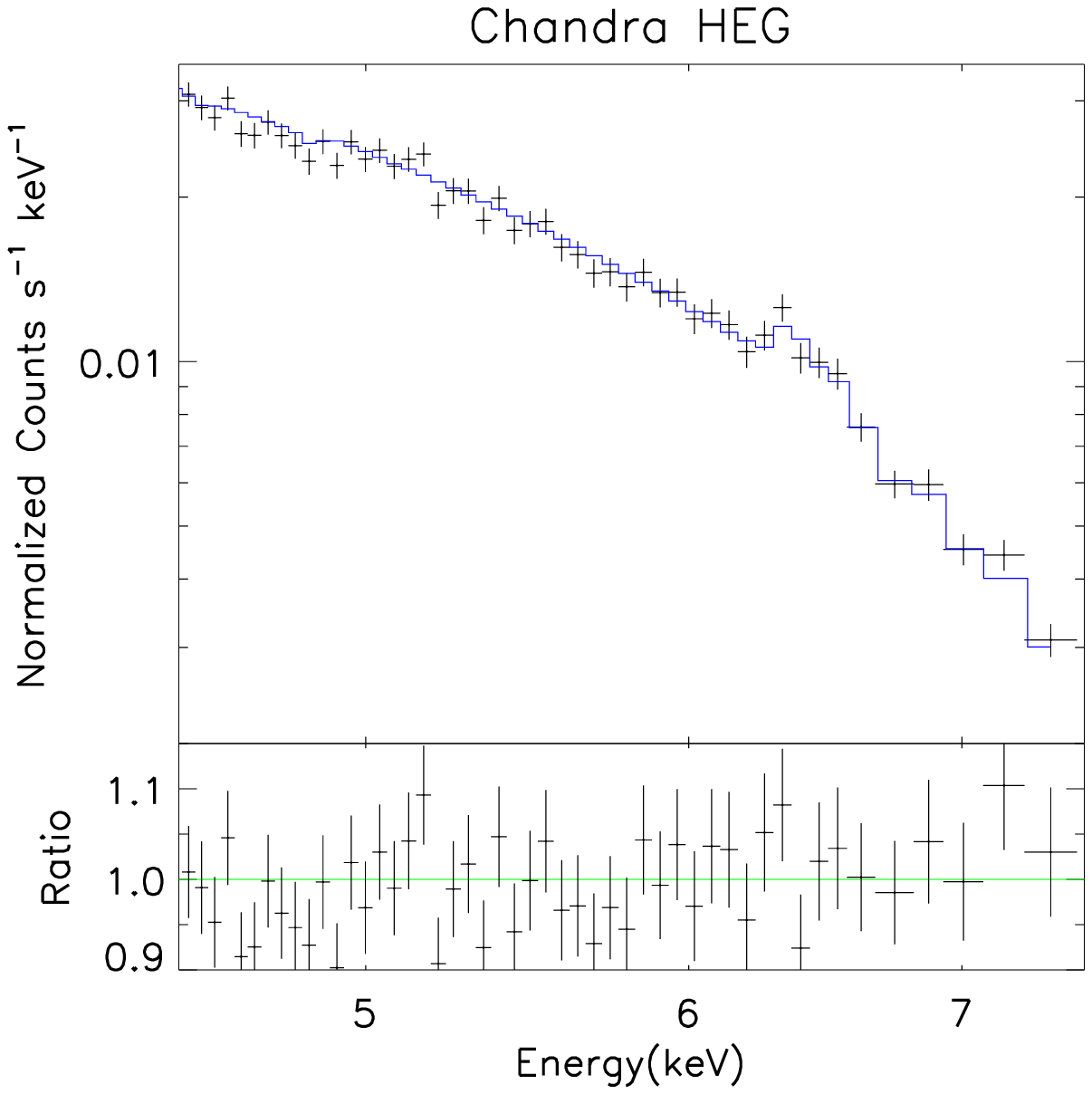}
                \end{center}
            \end{minipage}
            \begin{minipage}{0.33\textwidth}
                \begin{center}
                    \leavevmode \epsfxsize=50mm \epsfbox{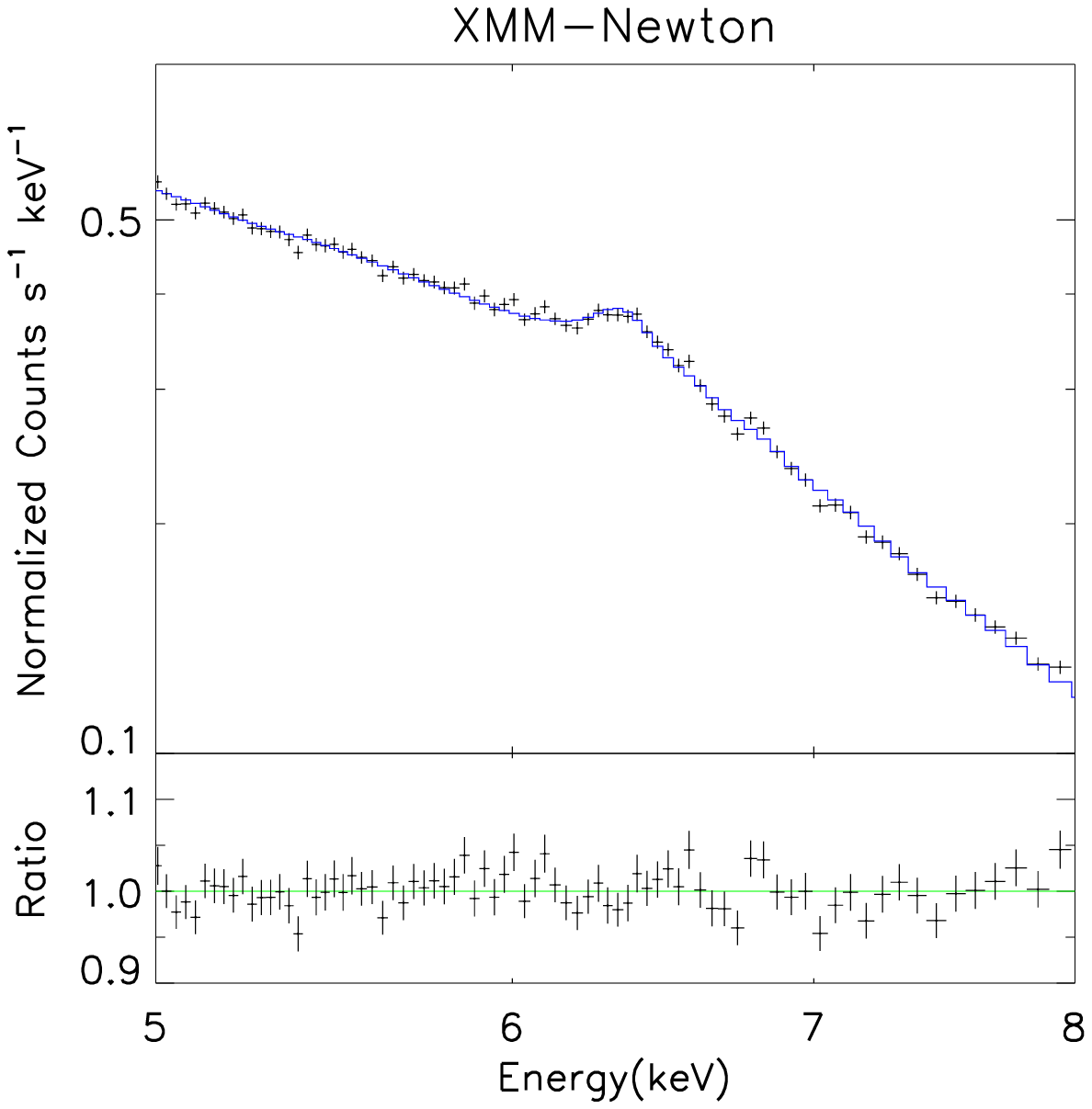}
                \end{center}
            \end{minipage}
            \begin{minipage}{0.33\textwidth}
                \begin{center}
                    \leavevmode \epsfxsize=50mm \epsfbox{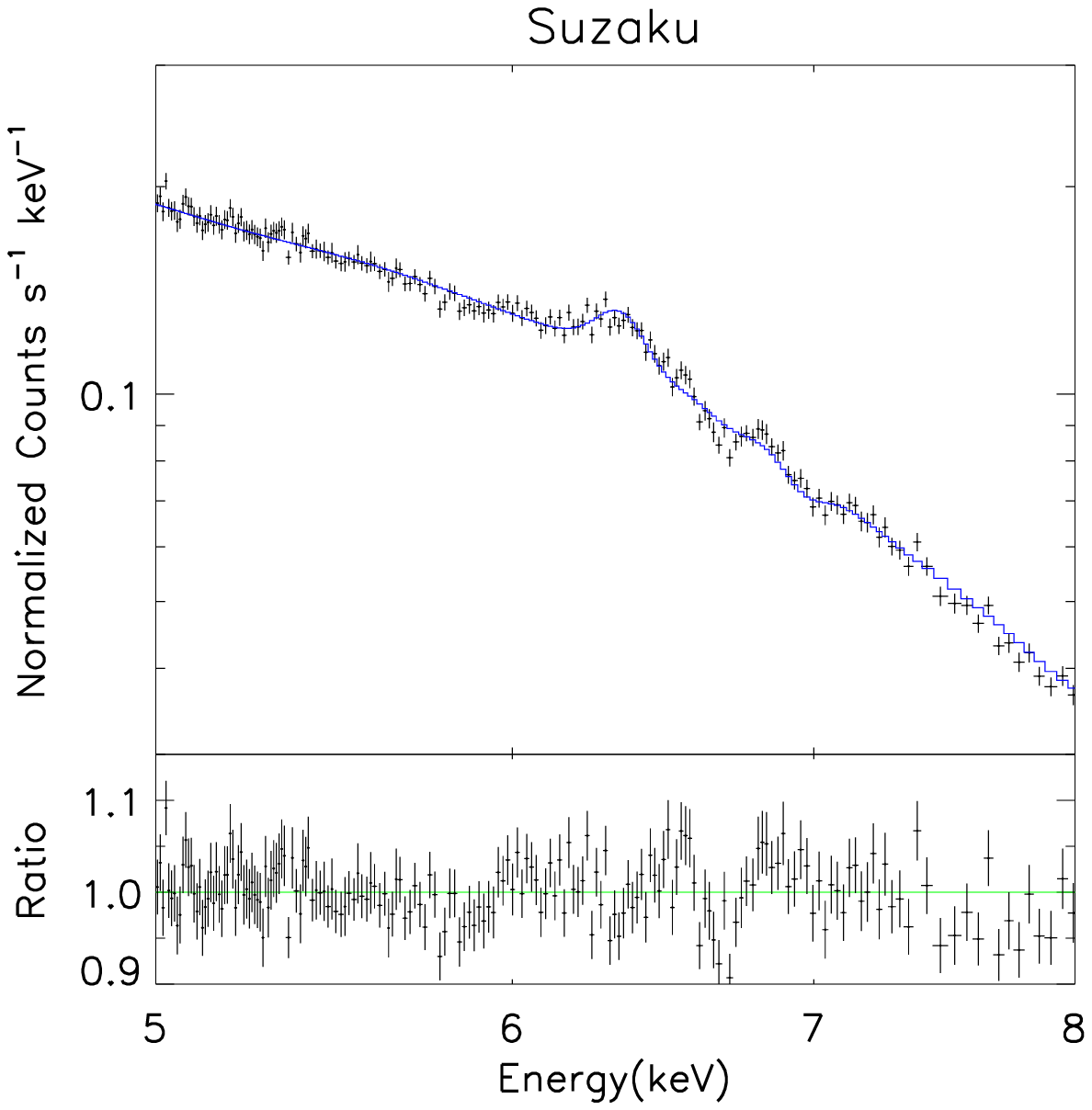}
                \end{center}
            \end{minipage}
            \caption{These figures show the zoom-in iron line fitted by the best-fitting model.
            Line profiles from left to right belong to \emph{Chandra} HEG , \emph{XMM} PN, and \emph{Suzaku} respectively. }
            \label{zoomiron}
        \end{minipage}
    \end{center}
\end{figure*}

We generated the ionised absorber models using {\sevensize XSTAR} {\sevensize V2.2.0}.
The source luminosity between 1 and 1000 Ryd was assumed to be $2\times10^{43}$ erg s$^{-1}$ \citep{Young05} with a
powerlaw spectrum shape at 0.1-20 keV, the gas density was set to be $10^{12}$ cm$^{-3}$, and the covering factor equal to 1, for all models.
Fitting parameters in these models are the photon index, the absorbing column, the ionisation parameter log $\xi$, and the iron and oxygen
abundances except for the fast component, whose oxygen abundance was fixed to unity.
Previous work adopted a photon index $\Gamma$ between 2.0 and 2.2 when calculating the absorption tables.
We allow $\Gamma$ of absorbers to vary with the value of the powerlaw incident continuum.
Iron and oxygen abundances are not fixed in case of super-solar or sub-solar abundant environment.
Turbulent velocities $v_{\mbox{\scriptsize turb}}$ = 100, 500, and 1000 km s$^{-1}$ and temperatures $T = 10^4$, $3 \times 10^4$, and
$10^5$ K were applied to produce different absorption models that cover a wide range of parameters.

We then tested these absorbers on the \emph{Chandra} MEG spectrum, with three absorbing zones and a blurred reflection component as a start.
Since the outflow velocities of both the slow and the fast components determined by previous work are in good agreement \citep[see][]{Holczer10},
we constrained the redshifts of the absorbing zones to vary in a small range.
The primary fitting results indicate a zone with $v_{\mbox{\scriptsize turb}}$ = 500 km s$^{-1}$ and two with
$v_{\mbox{\scriptsize turb}}$ = 100 km s$^{-1}$ are required to fit the spectrum, which is consistent with previous work.
The model was further tested on the low energy spectra ($< 3$ keV) of \emph{XMM} and \emph{Suzaku}, and we found the ionisation state of the fast
component is nearly constant on all datasets.
Therefore we fixed the ionisation parameter log $\xi$ to be 3.82 \citep{Miller07, Holczer10} in all the following fits.

A fourth absorbing zone, corresponding to the third kinetic system detected in \citet{Holczer10}, with
$v_{\mbox{\scriptsize turb}}$ = 100 km s$^{-1}$, $z = 0$, and $N_{\mbox{\scriptsize H}} = 4.06 \times 10^{20}$ cm$^{-2}$,
was added to test how much this component affects the fitting, and the detailed information of the absorbing zones we use is listed in
Table \ref{absorber}.
The results of other fitting parameters are shown in Table \ref{fitting}.
Clearly the model with the fourth component yields a better fit to the \emph{Chandra} spectrum of MCG-6-30-15.
The model struggles slightly below 1 keV, where the most residuals come from.
We then model the Fe L edge at 0.7 keV using dust in {\sevensize TBNEW}, and the result is shown in Fig. \ref{chandra}.

As to the difference spectra, we fit them with a model consisting of a powerlaw, Galactic absorption and the absorbing zones we use to fit the real spectrum.
The model fits all our difference spectra without difficulty,
indicating the reflection component is sufficiently removed by the difference spectrum technique.

\subsection{Full Spectrum}

When constructing a model for the full spectra, a relativistic blurred reflection component and a neutral reflection component should be
included in order to fit the skewed, broad iron line and a narrow iron line reflected from distant matter \citep[detected by][]{Young05}.
We model reflection from the innermost regions of the accretion disc with the self-consistent model
{\sevensize REFLIONX} \citep{Ross05} convolved by the {\sevensize KDBLUR} model, and the neutral reflection with
{\sevensize REFLIONX} only ($\xi = 1.0$).
The iron abundance $A_{\mbox{\scriptsize Fe}}$ of the reflection components are bound for consistency.
We set the outer radius of the accretion disc to be 400 R$_{\mbox{\scriptsize g}}$ in all fits, where R$_{\mbox{\scriptsize g}}$ is the gravitational radius GM/c$^{2}$.
Nevertheless, the two-reflection model is still not enough to fit the iron line structure perfectly, and we introduce another narrow line modelled by {\sevensize ZGAUSS}.
The new model works well on joint MEG and HEG spectra (see Fig. \ref{chandra}) and explains the complex structure of the iron line.

\begin{figure}
\begin{center}
\leavevmode \epsfxsize=8.5cm \epsfbox{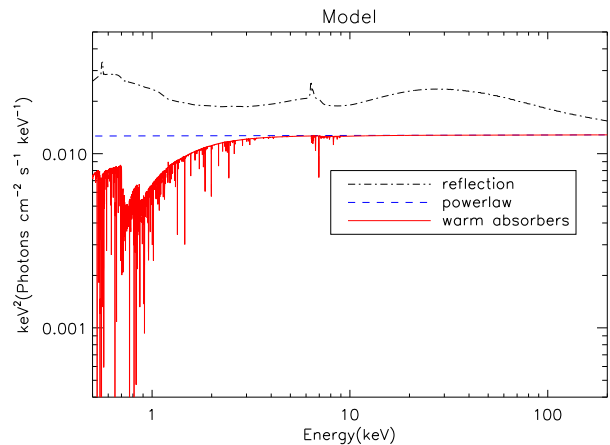}
\end{center}
\caption{The figure shows the decomposed components of the 4-absorption model. The black dot-dash line above shows the reflection component consisted of a blurred reflection, a neutral reflection and a narrow Gaussian line. The blue dash line represents a $\Gamma=2.0$ powerlaw and the red solid line shows the effects of the warm absorbers. Note that the reflection component has been moved above for clarity.}
\label{model}
\end{figure}

Some claimed the hard excess seen in some AGNs, included MCG-6-30-15, cannot be explained by the reflection model, and only clumpy absorbers model can work.
\citet{Miller08} first modelled the hard excess with absorbed reflection, but in their later work \citep[i.e.][]{Miller09} they claimed
the high energy emission is dominated by absorption, and the hard excess is reproduced by the continuum covered by a partially covering absorber.
To check the availability of the reflection model at high energies, we use the simultaneous \emph{XMM} and \emph{BeppoSAX} spectra and both \emph{Suzaku} XIS and PIN data,
adopting the same model used to fit the \emph{Chandra} HETGS spectra.
The results (see Fig. \ref{hard}) show the reflection model is able to produce a ``Compton hump" easily, and the hard excess can be well explained by simple reflection without any additional absorption.
The zoom-in iron lines of different datasets are shown in Fig. \ref{zoomiron}, and Fig. \ref{model} shows the decomposed components of the final model.

The resulting $\chi^{2}_{\nu}$ shown in Table \ref{fitting} may not be satisfactory enough.
If 1 percent systematic errors are included (using the {\sevensize SYSTEMATIC} command in {\sevensize XSPEC}) for the whole band of \emph{Suzaku}, and for \emph{XMM} and \emph{Chandra} above 1 keV, $\chi^{2}_{\nu}$ is acceptable at 1.00, 1.02 and 1.04, respectively.

\begin{table}
 \caption{The table lists parameters of the warm absorbers in the best-fitting model of different satellites,
 in which $N_{\mbox{\scriptsize H}}$ is given in $10^{21}$ cm$^{-2}$, temperature in Kelvin, and $\xi$ in erg cm s$^{-1}$.
 The column density of the local absorption component is fixed at the same value with Galactic absorption, as suggested in \citet{Holczer10},
 in all fittings.
 Meanwhile, the ionisation state of the fast component is fixed as well.}
\label{absorber}
\begin{tabular}{@{}clcccc}
\hline\hline
Absorber & fast & slow (1) & slow (2) & local\\
$T$ (K) & $3\times10^4$ & $10^4$ & $3\times10^4$ & $10^4$\\
\hline
\multicolumn{5}{c}{\emph{Chandra}}\\
\hline
$N_{\mbox{\scriptsize H}}$ & $209.4^{+36.9}_{-33.8}$ & $3.43^{+0.31}_{-0.42}$ & $0.27^{+0.20}_{-0.13}$ & (0.406)\\
log $\xi$ & (3.82) & $1.71\pm0.03$ & $2.47^{+0.03}_{-0.16}$ & $-1.62^{+0.05}_{-0.02}$ \\
\hline
\multicolumn{5}{c}{\emph{XMM} + \emph{BeppoSAX}}\\
\hline
$N_{\mbox{\scriptsize H}}$ & $27.4^{+9.5}_{-14.0}$ & $2.72^{+0.63}_{-0.28}$ & $0.99^{+0.46}_{-5.30}$ &  (0.406)\\
log $\xi$ & (3.82) & $1.68^{+0.05}_{-0.03}$ & $2.49^{+0.01}_{-0.10}$ & $-0.57\pm0.03$ \\
\hline
\multicolumn{5}{c}{\emph{Suzaku}}\\
\hline
$N_{\mbox{\scriptsize H}}$ & $38.9^{+10.4}_{-8.1}$ & $8.99^{+0.87}_{-0.91}$ & $0.14^{+0.12}_{-0.14}$ & (0.406)\\
log $\xi$ & (3.82) & $1.61^{+0.03}_{-0.04}$ & $1.73^{+0.11}_{-0.13}$ & $2.00^{+0.00}_{-0.13}$ \\
\hline\hline\\
\end{tabular}
\end{table}

\begin{table}
 \caption{The Table below lists the fitting parameters and $\chi^{2}$ obtained by the reflection model with different numbers of absorbing zones.
 The top block shows the result using a reflection model with four absorbers; the middle block displays the result fitted with a 3-absorber
 model (local absorption taken out from the 4-absorber model); the bottom block shows the result using a reflection model with one slow
 component and one fast component only. In the case of 2 warm absorbers, only \emph{XMM} and \emph{Suzaku} data have been tested.
 As mentioned in Section \ref{moretests}, emissivity profile index, iron abundance and inclination angle are fixed when testing \emph{XMM} and \emph{Suzaku} observations.}
\label{fitting}
\begin{tabular}{@{}cccc}
\hline\hline
parameter & \emph{Chandra} HETGS & \emph{XMM + BeppoSAX}  & \emph{Suzaku} \\
\hline
\multicolumn{4}{c}{model with 4 absorbing zones} \\
\hline
$\Gamma$ & $1.97\pm0.00$ & $2.00^{+0.00}_{-0.01}$ & $1.98\pm0.01$ \\
index & $8.00^{+0.00}_{-0.16}$ & $3.78^{+0.05}_{-0.08}$ & 3.09 \\
$R_{\mbox{\scriptsize in}}$(R$_{\mbox{\scriptsize g}}$) & $1.31^{+0.08}_{-0.00}$ & $1.57^{+0.13}_{-1.57}$ & 2.50 \\
$A_{\mbox{\scriptsize Fe}}$ & $1.79^{+0.10}_{-0.29}$ & $1.73^{+0.19}_{-0.12}$ & $4.00^{+0.00}_{-0.10}$ \\
$E_{\mbox{\scriptsize Fe}}$ & $6.53^{+0.06}_{-0.09}$ & $6.52\pm0.03$ & $6.38^{+0.01}_{-0.02}$ \\
$\phi$ & $35.0^{+0.6}_{-35.0}$$^{\circ}$ & $37.7^{+3.4}_{-2.2}$$^{\circ}$ & $44.0^{\circ}$ \\
$\chi^{2}/d.o.f.$ & 2417.7/2139 & 5059.3/3809 & 1684.7/1576 \\
\hline
\multicolumn{4}{c}{model with 3 absorbing zones} \\
\hline
$\Gamma$ & $1.94^{+0.01}_{-0.00}$ & $1.96\pm0.01$ & $1.98\pm0.01$ \\
index & $8.00^{+0.00}_{-0.10}$ & $3.84^{+0.09}_{-0.07}$ & 3.34 \\
$R_{\mbox{\scriptsize in}}$(R$_{\mbox{\scriptsize g}}$) & $1.31^{+0.04}_{-0.00}$ & $1.31^{+0.41}_{-1.31}$ & 2.66 \\
$A_{\mbox{\scriptsize Fe}}$ & $2.07^{+0.14}_{-0.13}$ & $2.77^{+0.79}_{-0.65}$ & $4.00^{+0.00}_{-0.11}$ \\
$E_{\mbox{\scriptsize Fe}}$ & $6.48^{+0.04}_{-0.07}$ & $6.49^{+0.04}_{-0.03}$ & $6.38^{+0.01}_{-0.02}$\\
$\phi$ & $35.7^{+1.4}_{-35.7}$$^{\circ}$ & $35.5^{+4.6}_{-3.1}$$^{\circ}$ & $41.6^{\circ}$ \\
$\chi^{2}/d.o.f.$ & 2516.4/2140 & 5111.9/3810 & 1688.3/1577 \\
\hline
\multicolumn{4}{c}{model with 2 absorbing zones} \\
\hline
$\Gamma$ & - & $1.98\pm0.01$ & $1.98\pm0.00$ \\
index & - & $4.40^{+0.12}_{-0.17}$ & 2.98 \\
$R_{\mbox{\scriptsize in}}$(R$_{\mbox{\scriptsize g}}$) & - & $1.24^{+0.33}_{-1.24}$ & 3.11 \\
$A_{\mbox{\scriptsize Fe}}$ & - & $1.00^{+0.01}_{-0.00}$ & $4.00^{+0.00}_{-0.17}$ \\
$E_{\mbox{\scriptsize Fe}}$ & - & $6.44^{+0.04}_{-0.05}$ & $6.37\pm0.02$ \\
$\phi$ & - & $40.6^{+3.8}_{-3.0}$$^{\circ}$ & $33.0^{\circ}$ \\
$\chi^{2}/d.o.f.$ & - & 6040.9/3815 & 1692.0/1582 \\
\hline\hline
\end{tabular}
\end{table}

\begin{table}
\caption{The table shows a comparison among different models.
\citet{Holczer10} and present work use full covering zones only, but
the former used a phenomenological cubic spline continuum, and the
later models the continuum with a powerlaw and a reflection. \citet{Miller09}
used both full and partial covering absorbers, and a simple continuum without reflection. $N_{\mbox{\scriptsize H}}$ is given in $10^{21}$ cm$^{-2}$, temperature in Kelvin, and $\xi$ in erg cm s$^{-1}$.}
\label{zones}
\begin{tabular}{@{}clccc}
\hline\hline
work & & Holczer et al. & Miller et al. & present work \\
\hline
\multicolumn{5}{c}{full covering zones}\\
\hline fast & $N_{\mbox{\scriptsize H}}$ & $81\pm7$ & (80.0) & $209.4^{+36.9}_{-33.8}$ \\
 & log $\xi$ & $3.82\pm0.03$ & (3.95) & (3.82) \\
slow(1) & $N_{\mbox{\scriptsize H}}$ & $2.3\pm0.3$ & $0.27\pm0.03$ & $3.43^{+0.31}_{-0.42}$ \\
 & log $\xi$ & -1.5-0.5 & $0.88\pm0.16$ & $1.71\pm0.03$ \\
slow(2) &  $N_{\mbox{\scriptsize H}}$ & $3.0\pm0.4$ & $11.8\pm0.5$ & $0.27^{+0.20}_{-0.13}$ \\
 & log $\xi$ & 1.5-3.5  & $2.39\pm0.01$ & $2.47^{+0.03}_{-0.16}$ \\
local &  $N_{\mbox{\scriptsize H}}$ & 0.40 & - & (0.406) \\
 & log $\xi$ & - & - & $-1.62^{+0.05}_{-0.02}$ \\
\hline
\multicolumn{5}{c}{partial covering zones}\\
\hline zone 1 & $N_{\mbox{\scriptsize H}}$ & - & $1910\pm300$ & - \\
 & log $\xi$ & - & - & -\\
zone 2 & $N_{\mbox{\scriptsize H}}$ & - & $29\pm1$ & - \\
 & log $\xi$ & - & $1.38\pm0.03$ & -\\
\hline\hline\\
\end{tabular}
\end{table}

\subsection{More Tests on \emph{XMM} and \emph{Suzaku}}
\label{moretests}

We test the 3-absorber model on \emph{XMM} and \emph{Suzaku} observations.
Since data points at low energies are statistically stronger than those at higher energies, the model tends to fit the low-energy spectrum and results in an unfitted iron line profile in \emph{Suzaku} data.
We then fit the spectrum above 3 keV first and fix the parameters of the convolution model (the emissivity profile index, inner radius $R_{\mbox{\scriptsize in}}$ and inclination angle $\phi$) so that the model will not ignore the iron line completely when fitting the full spectrum.
The results are shown in Table \ref{fitting}.

As stated in Section \ref{subsec:absorbers}, \emph{Chandra} HETGS data tend to include the local absorption component, and we test how necessary this component is needed on \emph{XMM} and \emph{Suzaku} observations as well.
The $\chi^{2}$ of the fittings of both \emph{XMM} \emph{Suzaku} data do improve (see Table \ref{fitting}), though not much in \emph{XMM} data.
However, the ionisation state of one of the slow sub-components in \emph{Suzaku} is close to the value of the local component (see Table \ref{absorber}), so it remains unclear whether \emph{Suzaku} data are able to resolve the additional absorption.
Whether the 4-absorber model or the 3-absorber model is used, the ionisation states of the slow component remain constant, indicating the local absorption may not much affect the fits.

The joint \emph{XMM} and \emph{BeppoSAX} and the \emph{Suzaku} observations do not show a strong tendency for the local absorption component.
To test if it is likely that even fewer absorbing zones, say, two as \citet{Miyakawa09} suggested, explain the spectrum, one of the slow sub-components was removed from the model.
The results of the 2-absorption model are also shown in Table \ref{fitting}.
The 2-absorption model results in a worse $\chi^{2}$ in both \emph{XMM} and \emph{Suszaku} datasets, implying at least three absorbing zones are required to fit the spectrum of MCG-6-30-15.

To examine if the real spectra in different periods have similarities like the difference spectra,  we put the best-fitting model of the joint \emph{XMM} and \emph{BeppoSAX} spectrum to the \emph{Suzaku} broadband spectrum and modify the normalisation slightly.
It seems the spectra of these observations above 3 keV look fairly similar, but there are obvious differences below 2 keV, where the absorption dominates the spectrum.
The fact that the iron line band and hard excess appear constant but variations in source luminosity below 10 keV motivated the work of the clumpy absorption model.
\citet{Miller09} suggested the long-term X-ray variability may be caused by variation in the covering factor.

Nevertheless, the differences below 2 keV in the spectra may be alternatively explained by changes in column densities \citep[see][]{Walton10} and ionisation states in warm absorbers.
Since there are outflowing components in MCG-6-30-15, column densities of these components may vary throughout time, and their ionisation states may possibly change due to the re-allocation of clouds.
We fit \emph{Suzaku} data by the best-fitting model obtained from the joint \emph{XMM} and \emph{BeppoSAX} spectrum, with all the parameters of warm absorbers fixed except the column densities.

%==============================================
\section{Discussion} \label{sec:discussion}
%==============================================

\subsection{Warm Absorbers}

Table \ref{fitting} shows the $\chi^{2}$ of the fits using the model
with and without the fourth local absorption component. Only
\emph{Chandra} data show a strong tendency to include the local
absorption component, which improves the fitting of \emph{Suzaku}
data but does not show strong evidence of existence, and \emph{XMM}
observation only gets a slightly better fit with it. The local absorption
component in fact contributes little to the spectrum, and only a few
oxygen lines around 20 {\AA} are slightly shifted by the component.
Besides, if this component is caused by the ISM in the local group as \citet{Holczer10} suggest, its ionisation state should stay constant in all datasets. But the ionisation states of the component obtained from \emph{Chandra} and \emph{Suzaku} are significantly different (see Table \ref{absorber}), indicating the data may have trouble constraining the parameters of this zone.

The result is expected because \emph{XMM} and \emph{Suzaku} do not have the sharp energy resolution of \emph{Chandra}.
\citet{Holczer10} claimed the third kinetic component was previously confused with the fast component because of similar velocities;
even \emph{Chandra} does not have enough resolution to constrain the line width.
Therefore it is not surprising that \emph{Suzaku} could not discern these two components,
considering the weak signature one of them.
Our spectral fitting results do favour its existence in the \emph{Chandra} spectra of MCG-6-30-15, but cannot judge whether the component is
caused by the Galactic absorption or another outflow; the issue is not relevant to the present work.
So far there have already been three kinetic systems, which are described by four ionisation absorbers in this work,
confirmed to be responsible for the soft X-ray band.
The highly ionised fast component, with $v_{\mbox{\scriptsize turb}}
=$ 500 km s$^{-1}$, is required in all the observations, though the
column density needed for each observation differs (see Table
\ref{absorber}). The fast component was reported in all previous
work of kinetic analysis of MCG-6-30-15, and our result again shows
its importance in modelling the spectra. The main features it causes
are the Fe {\sevensize XXV} ($\sim$ 6.7 keV) and Fe {\sevensize
XXVI} ($\sim$6.9 keV) absorption lines, which can be seen in Fig.
\ref{zoomiron}. The fast component is also the most important
absorber that affects the iron line band.

%The ionisation states of the slow component obtained by fitting with
%the 4-absorber model vary considerably among the observations.
Summarising the fitting results obtained by the 3-absorption and
2-absorption models (see Table \ref{fitting}), it seems \emph{XMM}
and \emph{Suzaku} data are able to distinguish the sub-components in
the slow outflow. Nonetheless, the values of ionisation states in
\emph{XMM} and \emph{Suzaku} data are different from that in
\emph{Chandra} data (see Table \ref{absorber}). This could be
because changes in ionisation states are needed to interpret the
variation in the low spectrum, or simply because \emph{XMM} and
\emph{Suzaku} spectra are not good enough to constrain parameters of
the sub-components well, though they do need two ionised zones to
model the slow component.

\subsection{Robustness of the Reflection Model}

The previous work \citep[e.g.][ etc.]{Vaughan04, Miniutti07} which
applied a reflection model on MCG-6-30-15 suggested the photon index
to be $\Gamma \sim 2-2.2$; the emissivity index is around 3; the
inner radius of the disc lies at somewhere less than 2
R$_{\mbox{\scriptsize g}}$; an iron abundance of 2 to 3 solar is needed to interpret the spectra, and the inclination
angle should be around $40^{\circ}$.
Most of the fitting parameters
of the relativistic reflection model, shown in Table \ref{fitting},
are in good agreement with the assumption of previous work. Note that the previous
work analysed the spectra above 3 keV only, and our work includes
the low energy band that is seriously influenced by the warm
absorbers. This implies the results obtained from the reflection
model would not be affected by absorbing zones and are reliable.

\citet{Zycki10} claimed the results received through the reflection
model are model-dependent. They found quite different powerlaw
indexes when they fit the \emph{Suzaku} spectrum above 3 keV
($\Gamma > 2.1$) and the spectrum above 1 keV ($\Gamma < 2.0$), but
we do not find this with our analysis. The $\Gamma$ we obtained from
a 3-absorber model or a 4-absorber model are all between 1.9 and
2.1, and the energy band we use is down to 0.5 keV. However, photon
index $\Gamma$ does change slightly when we adopt models with fewer
warm absorbers. The disagreement
might be caused by the warm absorbers used to model the soft X-ray
spectrum. The diverse photon indexes \citet{Zycki10} obtain is
simply due to the warm absorbers rather than the {\sevensize
REFLIONX} model. Additionally, the extra neutral ($\xi$ = 1.0)
{\sevensize REFLIONX} component that we used to fit the narrow iron
line works well, implying {\sevensize REFLIONX} can produce neutral
or mild ionised reflection without difficulty (we use a {\sevensize
REFLIONX} grid which extends to $\xi$ = 1.0, contrary to that used
by \citealt{Zycki10}).

\subsection{Comparisons with Previous Research}

Table \ref{zones} lists parameters of warm absorbers obtained by
different models. \citet{Holczer10} suggested the slow component
spans a considerable range of ionisation, which are -1.5 $<$ log
$\xi$ $<$ 0.5 and 1.5 $<$ log $\xi$ $<$ 3.5 for the sub-components, respectively. The ionisation state
of slow (1) in \citet{Miller09} and our best-fitting model for
\emph{Chandra} data (see Table \ref{zones}) do not fall in the
former range. The ionisation states of slow (2) of \citet{Miller09} and
present work are quite close and lie well in the expected range of
\citet{Holczer10}.

The column density we get for each absorbing zone differs significantly from
the values obtained in previous literature. They are probably the
most difficult parameters to be constrained, because results depend
on the way of modelling the continuum and the research tools that
have been used. Column densities are also sometimes under- or
over-estimated in order to fit complex structures in a spectrum as
complicated as in MCG-6-30-15, as discussed by \citet{Holczer10}.
The different parameters we used for generating {\sevensize XSTAR}
grids might again be the reason we have disagreement with previous
estimations. Moreover, {\sevensize REFLIONX} contributes some soft
emission and warm absorbers may need higher column densities to fit
the spectrum.

\citet{Miller08,Miller09} used two additional partial-covering
clumpy absorbing zones to model the spectrum. One of them mimicked
the shape of broadened asymmetric iron line, and the other partially
covered the continuum in order to explain the hard excess. The main
difference between the absorption-dominated model and the
reflection-dominated model is that the former has no distortions due to relativistic effects. \citet{Miller09} claimed the reflection
(relativistic effects not considered) model failed to interpret the
hard excess. In fact the hard excess can be simply explained by a
relativistically blurred reflection component without any
absorption. The complex iron line structure can also be fitted by a
combination of blurred and neutral reflections and absorptions
caused by a highly ionised warm absorber. Both the absorption and
the reflection models explain the spectrum and variability of
MCG-6-30-15. Our result does not rule out the ``3+2" model but shows
the reflection model that includes relativistic effects works. We consider this to be a more consistent physical scenario. Since the energy release in both models occurs very close to the black hole, where relativistic reflection is expected.

%==============================================
\section{Conclusion} \label{sec:conclusion}
%==============================================

The complex X-ray spectrum of MCG-6-30-15 can be well explained by a model consisting of relativistic reflection and three or four completely covered absorbing zones, in contrast with the conclusion of \citet{Miller08,Miller09}.
Our result does not rule out the absorption-dominated scenario but does support the reflection-dominated model.
In addition, we use absorbing zones already confirmed by previous analysis, while \citet{Miller08,Miller09} added two offset components to yield good fits.
The ``3+2" model is supported by the Principal Components Analysis \citep{Miller07}.
However, if there is an absorber responsible for the variability and the ``red wing" structure at 2-6 keV, structures such as curvature should appear in the difference spectra at this band, and it should leave some signature that could be traced by kinetic analysis.
The difference spectra retrieved from the observations we use (see Fig. \ref{diff}) is well fitted by a simple powerlaw above 3 keV, contrary to the assumption that the spectrum above 3 keV is dominated by absorption.

A highly-ionised fast outflow is confirmed by all observations. This is the most constrained absorbing zone among the warm absorbers needed to explain the X-ray spectrum of MCG-6-30-15.
The slow component with two ionisation states are confirmed as well, but \emph{XMM} and \emph{Suzaku} data are not good enough to determine the ionisation states correctly.
The ''local absorption" component detected by \citet{Holczer10} does improve our fitting to the \emph{Chandra} HETGS spectra.
By summarising the results we find three absorbing zones are required to fit \emph{XMM} and \emph{Suzaku} data, and a fourth zone is need to fit \emph{Chandra} data completely.
The 2-absorber model suggested by \citet{Miyakawa09} is not sufficient to interpret the complex spectrum of MCG-6-30-15, so the discrepancy in photon index between the spectrum above 3 keV and above 1 keV claimed by \citet{Zycki10} is due to the use of inappropriate warm absorbers.

We also test the robustness of our model and find the results are in good agreement with previous work.
Therefore we conclude that the reflection model can robustly interpret the spectra of MCG-6-30-15 in the range from 0.5-200 keV without any partial-covering absorbers.

\section*{Acknowledgements}
We thank Randy Ross for the extensive {\sevensize REFLIONX} grids used and Tim Kallman for providing {\sevensize XSTAR}.
CYC thank Dominic Walton for help in producing the {\sevensize XSTAR} grids.

\bibliographystyle{mn2e_uw}
\bibliography{mcg}

%\begin{thebibliography}{99}
%\end{thebibliography}

\label{lastpage}
\end{document}